\documentclass[conference]{IEEEtran}
\usepackage{amsmath,amsfonts,amssymb,bm}
\usepackage{algorithm}
\usepackage{algorithmic}
\usepackage{array}
\usepackage{booktabs}
\usepackage{cite}
\usepackage{graphicx}
\usepackage{subcaption}
\usepackage{textcomp}
\usepackage{url}
\usepackage{multirow}

\begin{document}

\title{Graph-Enhanced LLM for SWAN-ISAC}

\author{
\IEEEauthorblockN{Qian Gao, Ruikang Zhong, and Yuanwei Liu}
\IEEEauthorblockA{School of Electronic Engineering and Computer Science, Queen Mary University of London, London, U.K.\\
Email: q.gao@qmul.ac.uk}

\IEEEauthorblockA{School of Information and Communication Engineering, Xi’an Jiaotong University,  Xi’an 701149, P.R. China.\\
	Email: ruikang.zhong@xjtu.edu.cn}

\IEEEauthorblockA{Department of Electrical and Electronic Engineering, The University of Hong Kong, Hong Kong\\
Email: yuanwei@hku.hk}
}

\maketitle

\begin{abstract}
Segmented pinching antenna assisted integrated sensing and communication (ISAC) systems enable flexible spatial resource utilization by allowing different waveguide segments to be dynamically configured for transmission and reception. However, the resulting design requires the joint optimization of antenna deployment, segment partitioning, and beamforming under coupled communication and sensing constraints. In this paper, we propose a general learning framework for segmented pinching antenna assisted ISAC systems. Specifically, a channel state information (CSI)-induced self-graph is constructed to capture the scenario-dependent interactions among communication users and sensing targets. Based on the learned graph representation, a large language model (LLM) backbone with low-rank adaptation (LoRA) is employed, followed by two task-specific output heads for antenna deployment and beamforming prediction, respectively. Simulation results show that the proposed framework achieves a favorable tradeoff between communication rate and sensing accuracy.
\end{abstract}


\section{Introduction}
Integrated sensing and communication (ISAC) \cite{ISAC6G} is regarded as a key technology for future sixth-generation (6G) wireless networks, since it enables communication and sensing functionalities to share spectrum, hardware, and signal processing resources. By integrating these two functionalities, ISAC can improve spectral efficiency, hardware utilization, and environmental awareness across applications such as UAV-enabled ISAC, vehicular sensing, and localization-oriented wireless systems \cite{uav,driving,app,doa}. However, the communication and sensing tasks are inherently coupled in the spatial domain, where limited spatial degrees of freedom (DoF) must be carefully allocated to simultaneously guarantee communication throughput and sensing accuracy.

The array architecture plays a fundamental role in determining the achievable spatial DoF of ISAC systems. In line-of-sight (LoS) dominated scenarios, conventional fixed arrays often have limited reconfigurability and may not fully exploit spatial resources. Related reconfigurable aperture paradigms, such as intelligent reflecting surfaces, fluid antennas, and movable antennas \cite{ris,fuild,move}, provide useful design flexibility but do not directly capture the segment-wise radiating and receiving operation enabled by SWAN. Recently, segmented waveguide-enabled pinching antenna (SWAN) architectures \cite{seg} have been proposed as a promising alternative. By dividing the waveguide into multiple independently controlled segments and employing pinching antennas as reconfigurable radiating elements, SWAN offers flexible and cost-effective LoS link establishment with high structural freedom \cite{segpinch}. In addition, each segmented waveguide together with the pinching antennas mounted on it \cite{PA} can be configured for transmission or reception, which provides new opportunities for spatial resource allocation in ISAC systems.

The resulting design is challenging for three reasons. Firstly, antenna deployment, segment-wise transmit/receive partitioning, and beamforming are strongly coupled, leading to a mixed discrete-continuous optimization problem. Secondly, in practical ISAC scenarios, the numbers of communication users and sensing targets may vary over time, which makes fixed-input optimization and learning methods difficult to generalize. Thirdly, communication utility and sensing accuracy must be optimized jointly, since sensing-oriented constraints directly affect feasible beamforming designs. These factors make conventional model-based optimization \cite{model1,selection1,selection2,selection3} difficult to scale, especially for segmented pinching antenna-assisted ISAC systems with dynamic user configurations.

Recently, graph-based learning and large language models (LLMs) have shown strong potential for structured wireless optimization \cite{graphpinch1,graphpinch2,llm1,llm2}. Their expressive modeling ability makes them attractive for complex wireless design problems with heterogeneous inputs and coupled decision variables. However, directly applying an LLM to physical-layer optimization is nontrivial, since wireless inputs are structured numerical signals rather than text, and the learned model must remain compatible with system constraints and physics-based objectives. This motivates the development of structure-aware LLM-based learning architectures for segmented pinching antenna-assisted ISAC systems.

In this paper, we propose a general learning framework for segmented pinching antenna-assisted ISAC systems. The proposed framework combines a channel state information (CSI)-induced self-graph representation with an LLM backbone using low-rank adaptation (LoRA)\cite{lora}. Specifically, a CSI-induced self-graph neural network (SGNN) is first constructed to capture the interaction structure among communication users and sensing targets through CSI similarity, thereby producing permutation-invariant scenario representations for variable user settings. The graph-enhanced representation is then processed by an LLM backbone, followed by two task-specific output heads for antenna deployment and beamforming prediction, respectively. In this way, the proposed framework jointly predicts antenna deployment, segment partitioning, and communication/sensing beamforming variables in a unified manner. The main contributions of this paper are summarized as follows:
\begin{itemize}
\item We formulate a joint design problem for SWAN-ISAC systems, where antenna deployment, segment-wise transmit/receive partitioning, and beamforming are jointly optimized under communication-rate and sensing-accuracy requirements. The formulation explicitly captures the hierarchical structural flexibility of the SWAN architecture.
\item We propose a general learning framework that integrates a CSI-induced SGNN and an LLM backbone with LoRA. The self-graph provides permutation-invariant CSI representations for variable user settings, while two task-specific output heads are used to jointly predict deployment and beamforming variables.
\item Simulation results demonstrate that the proposed framework achieves a favorable tradeoff between communication rate and sensing accuracy.
\end{itemize}

\section{System Model and Problem Formulation}
\begin{figure}[t]
\centering
\includegraphics[width=\linewidth]{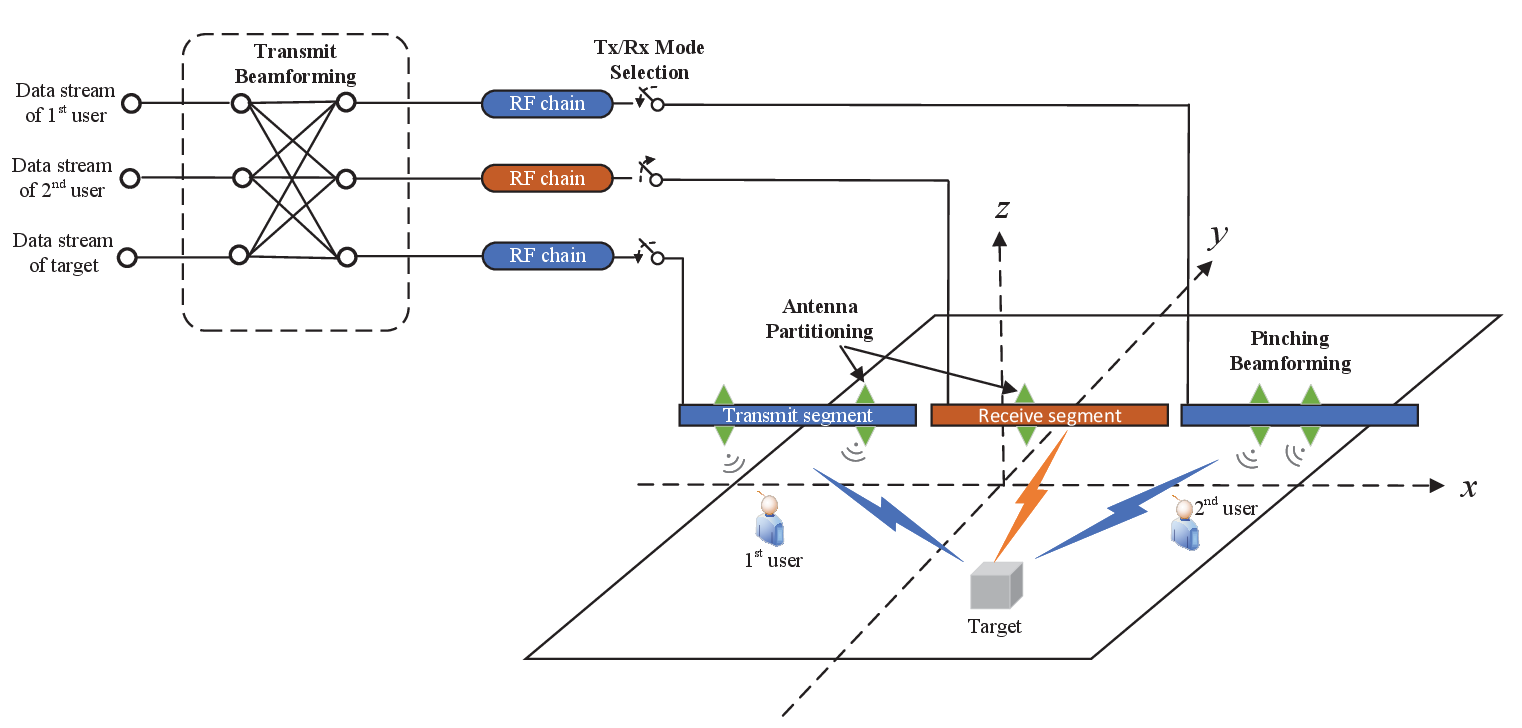}
\caption{SWAN-ISAC system with segment-wise Tx/Rx partitioning and flexible antenna deployment.}
\label{fig:system}
\end{figure}

We consider a SWAN-ISAC system, where a base station (BS) employs $M$ segmented dielectric waveguides and $N$ pinching antennas (PAs) to simultaneously serve $K_c$ communication users and sense $K_s$ targets, as illustrated in Fig.~\ref{fig:system}. The segmented waveguides are mounted at the BS side, and the PAs are deployed along the waveguide axis to form a reconfigurable aperture. Different from conventional fixed arrays, the SWAN architecture allows different waveguide segments to be assigned to transmission or reception, while the antenna deployment over the segmented structure jointly determines the communication and sensing performance.

For notational simplicity, we focus on one channel realization or one spatial scenario, and omit the time-slot index unless otherwise needed. The BS reference coordinates are denoted by $(x_{\mathrm{BS}},z_{\mathrm{BS}})$, and all PAs are deployed along the $y$-axis over a waveguide length $L$. The position of the $n$th PA is written as
\begin{equation}
\bm{\psi}_n=(x_{\mathrm{BS}}, y_n, z_{\mathrm{BS}}), \quad n\in\{1,\ldots,N\},
\end{equation}
where $y_n \in [0,L]$ denotes the deployment coordinate along the waveguide axis. The position vectors of the $k_c$th communication user and the $k_s$th sensing target are denoted by $\mathbf{p}_{k_c}\in\mathbb{R}^{3}$ and $\mathbf{p}_{k_s}\in\mathbb{R}^{3}$, respectively.

\subsection{Segment-Wise Transmit/Receive Partition}
Due to the segmented structure of SWAN, the first design layer is the segment-wise transmit/receive (Tx/Rx) partition. Let
\begin{equation}
\chi_m \in \{0,1\}, \quad m\in\{1,\ldots,M\},
\end{equation}
denote the operating mode of the $m$th segment, where $\chi_m=1$ and $\chi_m=0$ indicate that the segment is configured for transmission and reception, respectively. Accordingly, the transmit-segment set and receive-segment set are defined as
\begin{align}
\mathcal{M}_{\mathrm{t}} &\triangleq \{m \,|\, \chi_m=1\},\notag \\
\mathcal{M}_{\mathrm{r}} &\triangleq \{m \,|\, \chi_m=0\},
\end{align}
with $M_{\mathrm{t}} = |\mathcal{M}_{\mathrm{t}}|$ and $M_{\mathrm{r}} = |\mathcal{M}_{\mathrm{r}}|$.

To support multiuser communication while reserving a nonzero receiving aperture for sensing echo acquisition, the segment partition should satisfy
\begin{equation}
K_c +K_s\le M_{\mathrm{t}} \le M-1.
\end{equation}
This segment-wise partition determines which parts of the segmented aperture contribute to downlink transmission and which parts are reserved for sensing reception.

\subsection{Antenna Deployment over Segmented Waveguides}
The second design layer is the antenna deployment over the segmented waveguides. Let
\begin{equation}
\mathbf{y} = [y_1,\ldots,y_N]^T
\end{equation}
denote the deployment vector of the $N$ PAs along the waveguide axis. Since the waveguide length is $L$ and the structure is divided into $M$ equal segments, the $m$th segment corresponds to the interval
\begin{equation}
\mathcal{I}_m = \left[\frac{(m-1)L}{M}, \frac{mL}{M}\right), \quad m=1,\ldots,M-1,
\end{equation}
and
\begin{equation}
\mathcal{I}_M = \left[\frac{(M-1)L}{M}, L\right].
\end{equation}
Hence, the segment index associated with antenna $n$ is determined by the interval that contains $y_n$.

To ensure physical feasibility, the antenna deployment must satisfy the range constraint
\begin{equation}
0 \le y_n \le L, \quad \forall n,
\end{equation}
and the minimum-spacing constraint
\begin{equation}
|y_n-y_{n'}| \ge d_{\min}, \quad \forall n\neq n',
\end{equation}
where $d_{\min}$ is the minimum allowable inter-antenna spacing, typically chosen as half wavelength.

\subsection{Near-Field Signal and Channel Model}
We consider a near-field spherical-wave channel model \cite{nearfield}. For a node located at $\mathbf{p}\in\mathbb{R}^{3}$, the channel from the deployed PAs to this node is expressed as
\begin{equation}
\mathbf{h}(\mathbf{p})
=
\left[
\frac{\alpha e^{-j\frac{2\pi}{\lambda}\|\mathbf{p}-\bm{\psi}_1\|}}{\|\mathbf{p}-\bm{\psi}_1\|},
\ldots,
\frac{\alpha e^{-j\frac{2\pi}{\lambda}\|\mathbf{p}-\bm{\psi}_N\|}}{\|\mathbf{p}-\bm{\psi}_N\|}
\right]^T \in \mathbb{C}^{N\times 1},
\end{equation}
where $\alpha$ denotes the path-gain coefficient and $\lambda$ is the carrier wavelength. Accordingly, the channels toward the $k_c$th communication user and the $k_s$th sensing target are denoted by $\mathbf{h}_{k_c}\triangleq \mathbf{h}(\mathbf{p}_{k_c})$ and $\mathbf{h}_{k_s}\triangleq \mathbf{h}(\mathbf{p}_{k_s})$, respectively.

Let $\mathbf{w}_{k_c}\in\mathbb{C}^{N\times 1}$ denote the beamforming vector for the $k_c$th communication user and let $\mathbf{f}_{k_s}\in\mathbb{C}^{N\times 1}$ denote the probing beamforming vector for the $k_s$th sensing target. To reflect the segment-wise partition, the beamforming vectors are effectively constrained by the transmit-segment configuration, such that only antennas belonging to transmit-mode segments contribute to the transmitted signal. Let $\rho_c\in[0,1]$ and $\rho_s\in[0,1]$ denote the communication and sensing power-splitting factors, respectively, satisfying
\begin{equation}
\rho_c+\rho_s \le 1.
\end{equation}
Then the transmitted baseband signal can be written as
\begin{equation}
\mathbf{x}=
\sqrt{\rho_c}\sum_{k_c=1}^{K_c}\mathbf{w}_{k_c}s_{k_c}
+
\sqrt{\rho_s}\sum_{k_s=1}^{K_s}\mathbf{f}_{k_s}q_{k_s},
\end{equation}
where $s_{k_c}\sim\mathcal{CN}(0,1)$ is the information symbol for the $k_c$th communication user and $q_{k_s}\sim\mathcal{CN}(0,1)$ is the probing symbol associated with the $k_s$th sensing target.

\subsection{Communication and Sensing Metrics}
For communication, the received signal at communication user $k_c$ is given by
\begin{equation}
r_{k_c}
=
\mathbf{h}_{k_c}^T \mathbf{x} + n_{k_c},
\end{equation}
where $n_{k_c}\sim\mathcal{CN}(0,\sigma_c^2)$ is the additive white Gaussian noise. Therefore, the signal-to-interference-plus-noise ratio (SINR) of user $k_c$ is
\begin{equation}
\mathrm{SINR}_{k_c}
=
\frac{\rho_c |\mathbf{h}_{k_c}^T \mathbf{w}_{k_c}|^2}
{\rho_c \sum_{i\neq k_c} |\mathbf{h}_{k_c}^T \mathbf{w}_{i}|^2
	+\rho_s \sum_{j=1}^{K_s} |\mathbf{h}_{k_c}^T \mathbf{f}_{j}|^2
	+\sigma_c^2}.
\end{equation}
Accordingly, the achievable sum rate is expressed as
\begin{equation}
R_{\mathrm{sum}}=\sum_{k_c=1}^{K_c}\log_2(1+\mathrm{SINR}_{k_c}),
\end{equation}

For sensing, we adopt the position estimation accuracy of the sensing targets as the performance metric. Specifically, let $\mathbf{J}_{k_s}$ denote the Fisher information matrix (FIM) associated with the estimation of the $k_s$th target position. Then the corresponding Cram\'er--Rao lower bound (CRLB) matrix is given by $\mathbf{J}_{k_s}^{-1}$, and the sensing accuracy metric is defined as
\begin{equation}
\mathrm{CRLB}_{k_s}
\triangleq
\mathrm{tr}\!\left(\mathbf{J}_{k_s}^{-1}\right),
\end{equation}
where a smaller value indicates better sensing accuracy. The communication and sensing objectives are inherently coupled. Increasing communication power or steering beams more aggressively toward users may improve $R_{\mathrm{sum}}$, but it may also reduce the effective sensing aperture or degrade the estimation accuracy of the sensing targets. This tradeoff is further affected by the antenna deployment and segment-wise Tx/Rx partition.

\subsection{Problem Formulation}
Based on the above model, we aim to jointly optimize the antenna deployment vector $\mathbf{y}$, the segment-wise Tx/Rx partition vector $\boldsymbol{\chi}=[\chi_1,\ldots,\chi_M]^T$, and the communication/sensing beamforming variables $\{\mathbf{w}_{k_c}\}_{k_c=1}^{K_c}$ and $\{\mathbf{f}_{k_s}\}_{k_s=1}^{K_s}$ so as to maximize the communication utility while satisfying sensing and physical-feasibility requirements. The resulting problem can be formulated as
\begin{subequations}\label{eq:opt_problem_conf}
\begin{align}
\max_{\mathbf{y},\,\boldsymbol{\chi},\,\{\mathbf{w}_{k_c}\},\,\{\mathbf{f}_{k_s}\}}
&\quad R_{\mathrm{sum}} \notag \\
\text{s.t.}\quad
&\quad \mathrm{CRLB}_{k_s} \le \varepsilon_{k_s}, \quad \forall k_s, \\
&\quad 0 \le y_n \le L, \quad \forall n, \\
&\quad |y_n-y_{n'}| \ge d_{\min}, \quad \forall n\neq n', \\
&\quad \chi_m \in \{0,1\}, \quad \forall m, \\
&\quad K_c + K_s \le \sum_{m=1}^{M}\chi_m \le M-1, \\
&\quad \rho_c \sum_{k_c=1}^{K_c}\|\mathbf{w}_{k_c}\|_2^2
+\rho_s \sum_{k_s=1}^{K_s}\|\mathbf{f}_{k_s}\|_2^2
\le P_{\max}.
\end{align}
\end{subequations}
where $\varepsilon_{k_s}$ is the maximum tolerable sensing error threshold for target $k_s$, and $P_{\max}$ is the total transmit power budget.

Problem \eqref{eq:opt_problem_conf} is highly challenging due to the strong coupling among deployment, partitioning, and beamforming. In particular, the segment-wise Tx/Rx partition introduces discrete structural decisions, whereas the deployment and beamforming variables are continuous and jointly affect both communication rate and sensing CRLB. These observations motivate the learning-based structured design developed in the next section.

\section{Proposed LLM-Enabled Framework}
For each scenario, the input CSI tensor is denoted by $\mathbf{H}\in\mathbb{R}^{(K_c+K_s)\times N\times 2}$, where the last dimension corresponds to the real and imaginary parts. The proposed model jointly predicts antenna deployment $\hat{\mathbf{y}}$, segment-wise partition $\hat{\boldsymbol{\chi}}$, and beamforming variables $\hat{\mathbf{W}}$ and $\hat{\mathbf{F}}$. The overall design follows a graph-enhanced LLM pipeline composed of a CSI-induced self-graph encoder, a LoRA-adapted LLM backbone, and two task-specific output heads. Relative to using a plain ordered-input neural network, this design explicitly separates sample-dependent relational extraction, high-capacity shared representation learning, and task-oriented decoding.

\subsection{CSI-Induced Self-Graph}
To capture scenario-dependent relations, we build a graph $\mathcal{G}=(\mathcal{V},\mathcal{E},\mathbf{A})$ whose nodes correspond to communication users and sensing targets. For node $i$, the feature is
\begin{equation}
\mathbf{x}_i=
\left[
\|\mathbf{h}_i\|_2,\;
\angle\!\Big(\sum_{n=1}^{N}[\mathbf{h}_i]_n\Big),\;
\tau_i
\right]^T,
\end{equation}
where $\tau_i$ indicates whether the node represents a communication user or a sensing target. The edge weight between nodes $i$ and $j$ is defined as
\begin{equation}
A_{ij}=
\frac{|\mathbf{h}_i^H\mathbf{h}_j|}
{\|\mathbf{h}_i\|_2\|\mathbf{h}_j\|_2+\epsilon}.
\end{equation}
After row normalization, the adjacency matrix encodes CSI similarity and produces a permutation-invariant relational representation. This construction is particularly useful when communication users and sensing targets appear in different orders across samples, since the graph depends on the underlying interaction pattern rather than on the raw input ordering.

A multi-layer self-graph encoder then generates a global scenario embedding
\begin{equation}
\mathbf{z}_g=\frac{1}{K_c+K_s}\sum_{i=1}^{K_c+K_s}\mathbf{h}_i^{(L_g)},
\end{equation}
which summarizes the interaction structure among users and targets.

\subsection{LoRA-Adapted LLM Backbone and Split Heads}
The antenna-domain CSI is tokenized across the $N$ deployed positions and projected into a hidden sequence $\tilde{\mathbf{X}}$. The graph embedding is injected through additive conditioning,
\begin{equation}
\bar{\mathbf{x}}_n=\tilde{\mathbf{x}}_n+\mathbf{W}_g\mathbf{z}_g, \quad n=1,\ldots,N.
\end{equation}
The conditioned sequence is processed by a pretrained GPT-style backbone with LoRA modules inserted into the main projection layers. This design preserves most backbone parameters while enabling efficient adaptation. Compared with training a large model from scratch, LoRA reduces the number of trainable parameters and helps stabilize optimization under the limited-size synthetic dataset used in this work.

Two task-specific heads decode the shared hidden representation. The deployment head predicts the raw antenna positions and segment logits,
\begin{equation}
[\hat{\mathbf{y}}^{\mathrm{raw}},\bm{\pi}_\chi]=g_{\mathrm{dep}}(\mathbf{H}_{\mathrm{LLM}}),
\end{equation}
and the final deployment is obtained as $\hat{\mathbf{y}}=L\cdot \sigma(\hat{\mathbf{y}}^{\mathrm{raw}})$. The segment-wise partition is produced from the corresponding logits and captures the transmit/receive mode of each segment. This design is more compact than directly predicting a full element-wise binary activation mask and better matches the structural characteristics of SWAN, where the key structural decisions are the antenna deployment and the segment-wise Tx/Rx partition.

In parallel with the deployment head, the beamforming head predicts the real and imaginary parts of the communication and sensing beamformers. The resulting outputs are combined into the complex beamforming matrices $\hat{\mathbf{W}}$ and $\hat{\mathbf{F}}$, which are then evaluated together with the predicted deployment and partition in the differentiable SWAN-ISAC environment. Therefore, the beamforming head is trained jointly with the deployment head under the same communication-sensing objective. Decoupling the two heads helps separate site-dependent structural learning from task-dependent beam synthesis. In practice, this decomposition is beneficial because the structural variables and the beamforming variables influence the final ISAC objective in different ways and may therefore benefit from partially specialized decoding paths.

\subsection{Training Objective}
The training loss combines deployment supervision, communication--sensing performance optimization, and geometry regularization:
\begin{equation}
\mathcal{L}=w_{\mathrm{dep}}\mathcal{L}_{\mathrm{dep}}+\mathcal{L}_{\mathrm{perf}}+\mathcal{L}_{\mathrm{geom}}.
\end{equation}
Here, $\mathcal{L}_{\mathrm{dep}}$ supervises the sorted deployment vector, while $\mathcal{L}_{\mathrm{perf}}$ maximizes $R_{\mathrm{sum}}$ and penalizes CRLB violation. The regularizer $\mathcal{L}_{\mathrm{geom}}$ enforces spacing and segment-feasibility consistency.
The joint loss is important because it prevents the network from optimizing one aspect of the problem in isolation. In particular, deployment supervision anchors the structural prediction, while the communication and sensing terms ensure that the predicted aperture configuration remains useful for the end performance metrics rather than merely matching geometric labels.

\begin{algorithm}[htbp]
\caption{Training Procedure of the Proposed Framework}
\label{alg:training_framework_short}
\begin{algorithmic}[1]
\REQUIRE Training dataset $\mathcal{D}$, system configuration $\mathcal{C}$.
\ENSURE Trained model parameters.
\STATE Initialize the CSI-induced self-graph encoder, the shared LLM backbone with LoRA adaptation, the deployment head, and the beamforming head.
\FOR{each training epoch}
\FOR{each mini-batch $(\mathbf{H}, \mathbf{y}^{\star}, \{\mathbf{p}_{k_c}\}, \{\mathbf{p}_{k_s}\}) \in \mathcal{D}$}
\STATE Construct the CSI-induced self-graph and obtain the graph embedding $\mathbf{z}_g$.
\STATE Tokenize $\mathbf{H}$, inject $\mathbf{z}_g$, and obtain the shared hidden representation $\mathbf{H}_{\mathrm{LLM}}$.
\STATE Predict $\hat{\mathbf{y}}$ and $\hat{\boldsymbol{\chi}}$ using the deployment head.
\STATE Predict $\hat{\mathbf{W}}$ and $\hat{\mathbf{F}}$ using the beamforming head.
\STATE Evaluate $R_{\mathrm{sum}}$ and $\mathrm{CRLB}$ in the differentiable SWAN-ISAC environment.
\STATE Compute the total loss $\mathcal{L}=w_{\mathrm{dep}}\mathcal{L}_{\mathrm{dep}}+\mathcal{L}_{\mathrm{perf}}+\mathcal{L}_{\mathrm{geom}}$.
\STATE Update the trainable parameters by backpropagation.
\ENDFOR
\ENDFOR
\end{algorithmic}
\end{algorithm}

\section{Simulation Results}
\begin{figure*}[!t]
\centering
\begin{subfigure}[b]{0.47\textwidth}
\centering
\includegraphics[width=\linewidth]{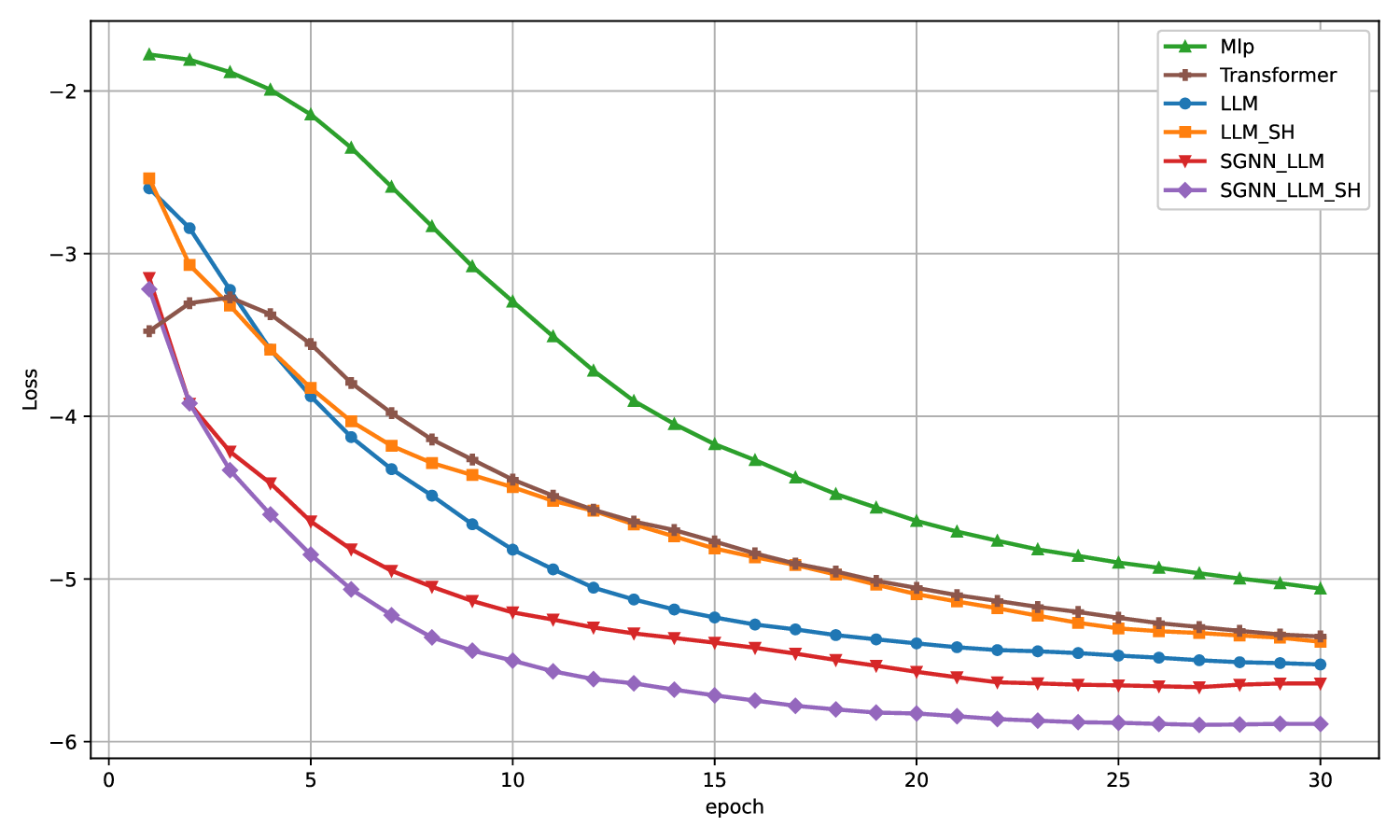}
\caption{Loss}
\label{fig:loss}
\end{subfigure}
\hfill
\begin{subfigure}[b]{0.47\textwidth}
\centering
\includegraphics[width=\linewidth]{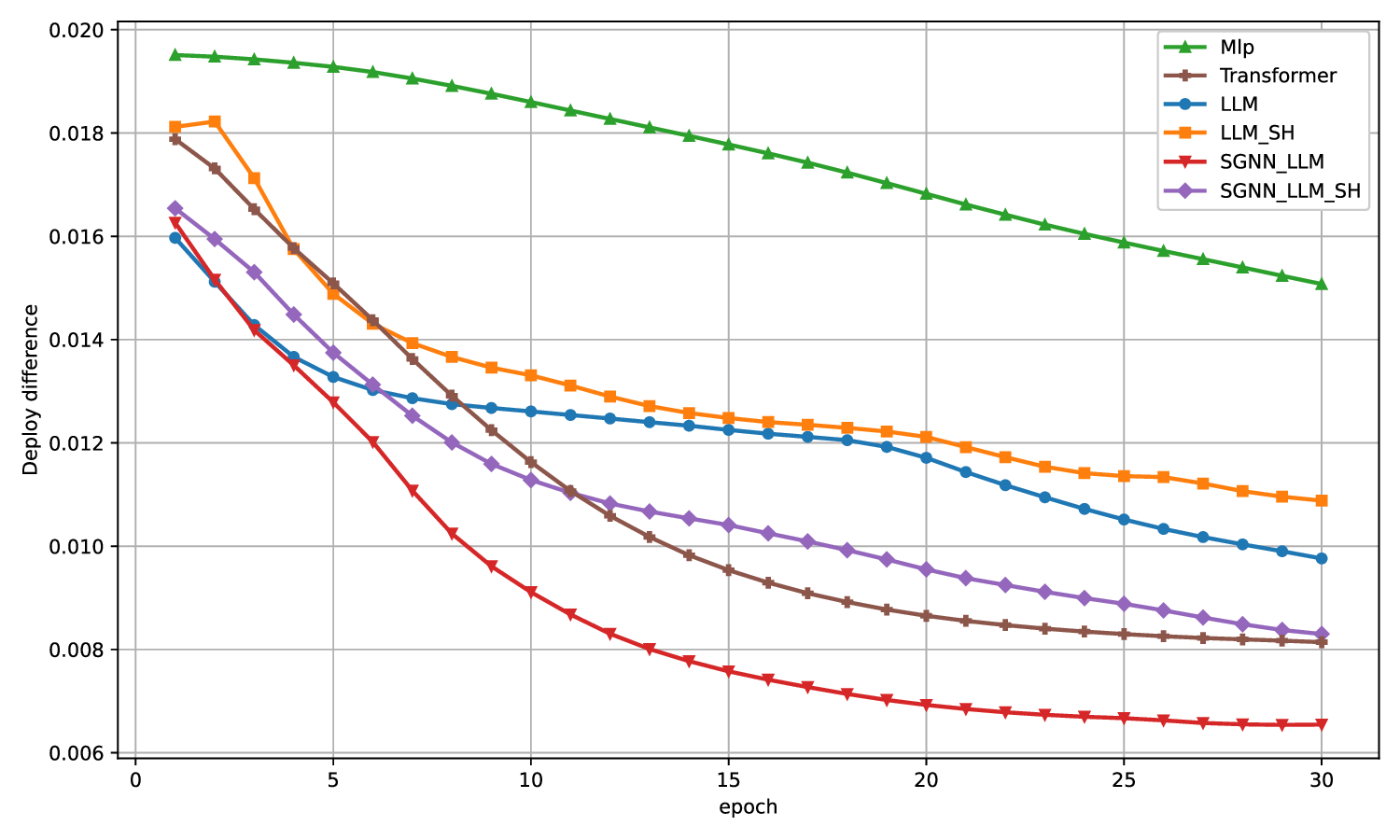}
\caption{Deployment diff.}
\label{fig:deploy}
\end{subfigure}

\vspace{0.2cm}

\begin{subfigure}[b]{0.47\textwidth}
\centering
\includegraphics[width=\linewidth]{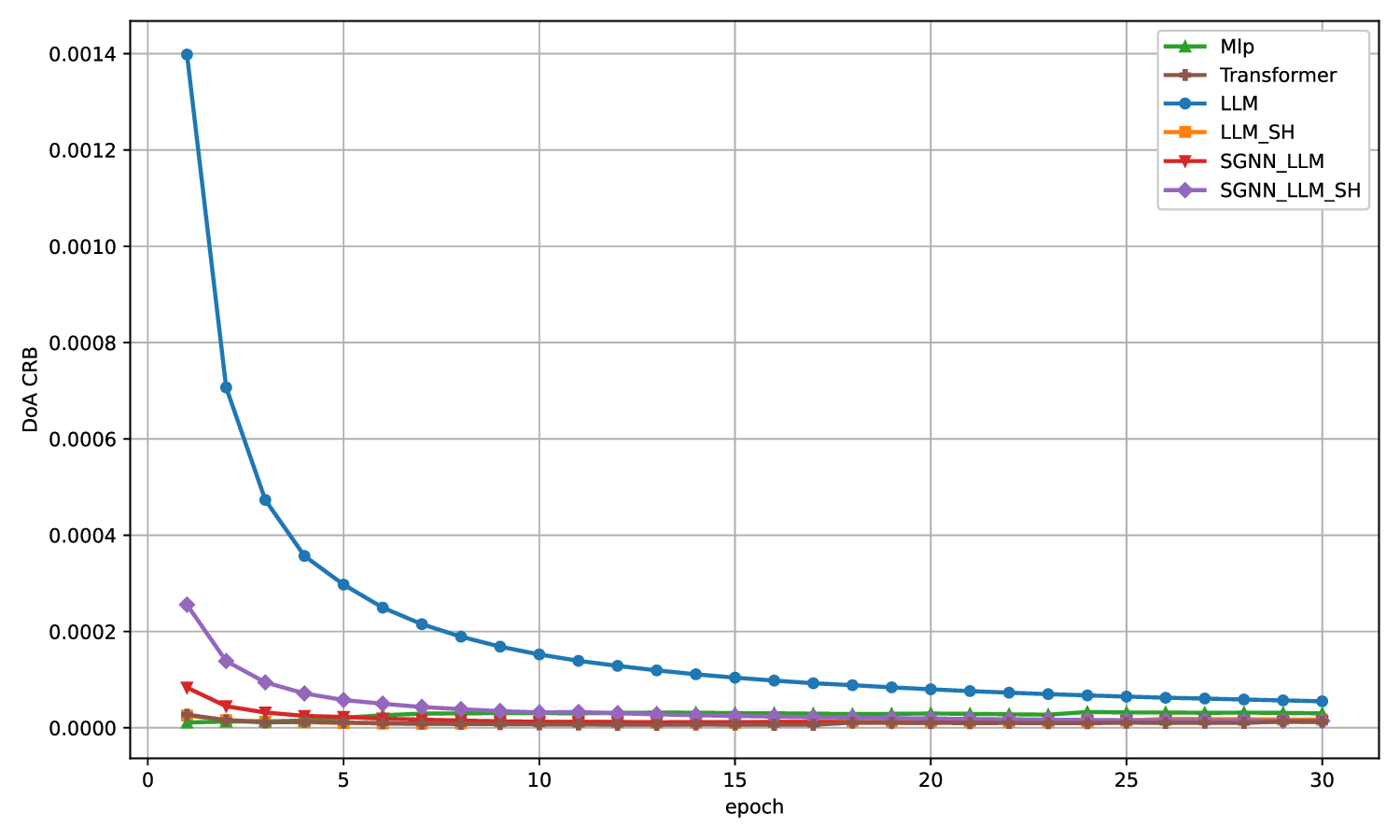}
\caption{CRLB}
\label{fig:doa}
\end{subfigure}
\hfill
\begin{subfigure}[b]{0.47\textwidth}
\centering
\includegraphics[width=\linewidth]{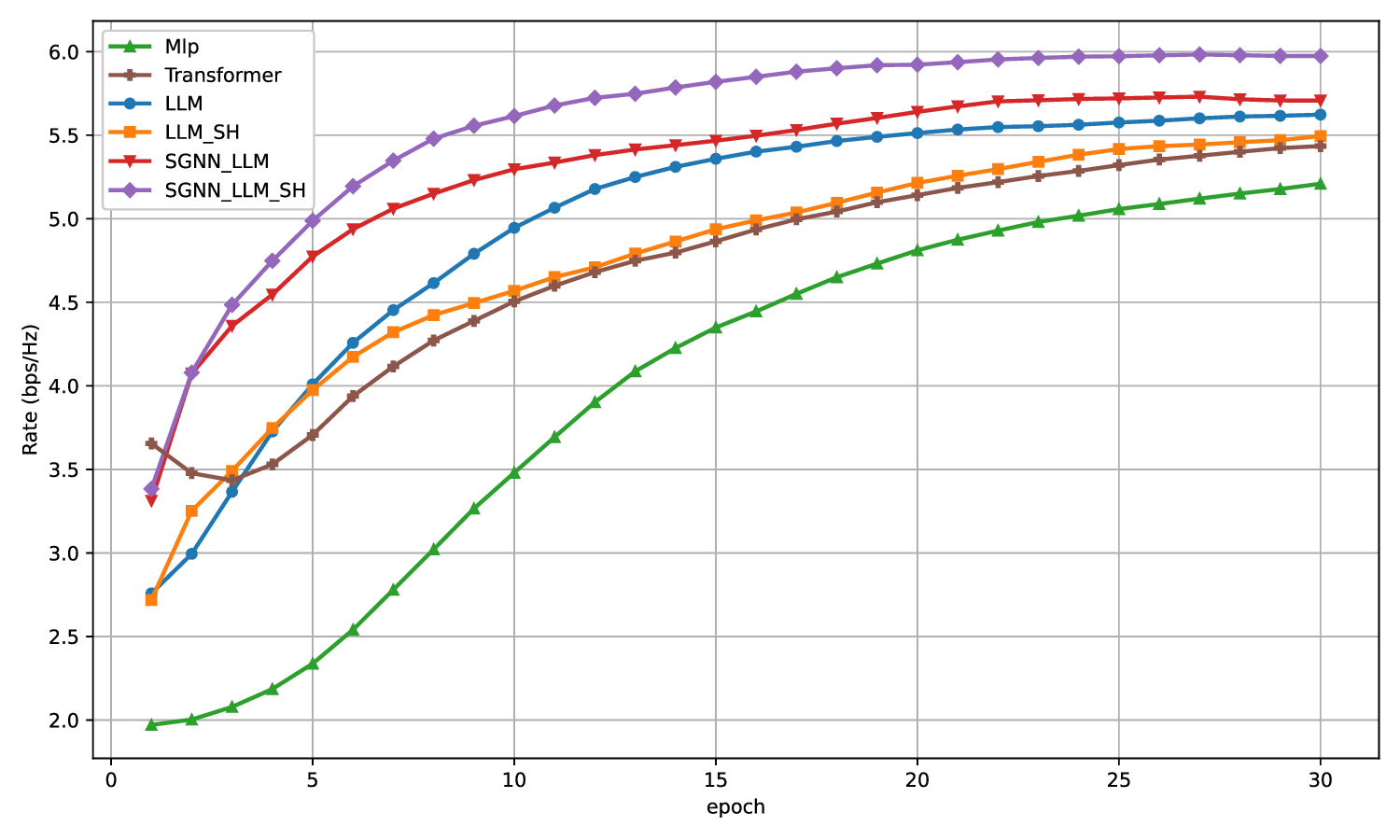}
\caption{Sum rate}
\label{fig:rate}
\end{subfigure}
\caption{Convergence behavior and benchmark comparison.}
\label{fig:benchmark}
\end{figure*}

Unless otherwise specified, the benchmark experiments use $M=4$ segmented waveguides, $N=40$ pinching antennas, $K_c=2$ communication users, $K_s=1$ sensing target, waveguide length $L=50$ m, and transmit power budget $P_{\max}=10$. We generate 3000 near-field channel realizations and split them into training, validation, and test sets with a ratio of 70\%/15\%/15\%. All methods are trained for 30 epochs using Adam with learning rate $5\times 10^{-4}$, training batch size $64$, and evaluation batch size $128$. Gradient clipping and exponential moving average evaluation are used for stability. The default loss weights are $w_{\mathrm{rate}}=1.0$, $w_{\mathrm{CRLB}}=0.2$, and $w_{\mathrm{dep}}=10.0$. For the LLM-based methods, the LoRA hyperparameters are rank $r=32$, scaling factor $\alpha=16$, and dropout $0.05$.

The compared baselines include MLP, Transformer, LLM, LLM\_SH, SGNN\_LLM, and SGNN\_LLM\_SH. Here, MLP directly regresses from flattened CSI, Transformer models antenna-domain CSI as a sequence, LLM uses a LoRA-adapted GPT backbone, LLM\_SH adopts split output heads, SGNN\_LLM prepends the CSI-induced self-graph encoder, and SGNN\_LLM\_SH denotes the proposed model. All methods are trained under the same data split and optimization protocol for a fair comparison.

Fig.~\ref{fig:benchmark} compares the validation loss, deployment error, sensing CRLB, and achievable rate of all methods under the same training protocol. Several observations can be made. First, SGNN\_LLM\_SH achieves the lowest validation loss and converges more steadily than the baseline methods, which indicates that the proposed graph-enhanced LLM architecture is better suited to the coupled SWAN-ISAC design. Second, in terms of communication performance, SGNN\_LLM\_SH consistently attains the highest achievable rate, while SGNN\_LLM and the LLM-based baselines form the second-best group. This suggests that the self-graph improves the feature quality provided to the LLM backbone, and that the decoupled prediction heads help the model better allocate representation capacity across structural and beamforming subtasks.

Third, the sensing and deployment curves show that purely ordered-input models such as MLP and Transformer are less effective in maintaining a balanced structural and sensing performance. In particular, the MLP baseline converges slowly and remains consistently inferior in both deployment error and achievable rate, which implies that simple flatten-and-regress mappings are inadequate for the hierarchical SWAN design space. The Transformer baseline improves over MLP, but it still underperforms the LLM-based methods, indicating that a moderate-size encoder is not sufficient to capture the highly coupled relations among deployment, partitioning, and beamforming variables.

Fourth, although SGNN\_LLM is highly competitive in deployment prediction, the split-head design in SGNN\_LLM\_SH yields the best overall communication--sensing tradeoff. This difference is meaningful: a model that predicts slightly more accurate deployment coordinates does not necessarily produce the best end-to-end ISAC performance, because communication utility and sensing accuracy depend on the joint interaction among all predicted variables. The results therefore support evaluating SWAN-ISAC learning methods through multiple coupled metrics rather than through one structural metric alone.

Finally, the CRLB curves indicate that the proposed model maintains competitive sensing behavior while improving rate. This is important because the rate improvement is not obtained by simply neglecting the sensing requirement. Instead, the benchmark suggests that the proposed graph-enhanced LLM is better able to learn a feasible region that preserves sensing quality while extracting more communication gain from the segmented aperture. Overall, the main benchmark confirms the effectiveness of combining CSI-induced relational modeling, parameter-efficient LLM adaptation, and task-specific decoding for segmented pinching antenna assisted ISAC.

\section{Conclusion}
This paper studied the joint optimization of antenna deployment, segment partitioning, and beamforming in segmented pinching antenna assisted ISAC systems. We proposed a general learning framework that combines a CSI-induced self-graph encoder with an LLM backbone using LoRA adaptation, followed by two task-specific output heads for deployment and beamforming prediction, respectively. Simulation results demonstrated that the proposed framework achieves a favorable tradeoff between communication rate and sensing accuracy compared with representative baselines.

\appendices
\section{Simplified Derivation of the Position CRLB}
Consider one sensing target with position parameter vector $\boldsymbol{\eta}=[x,y,z]^T$ and steering response $\mathbf{a}(\boldsymbol{\eta})\in\mathbb{C}^{N\times 1}$. Let $\mathbf{u}\in\mathbb{C}^{N\times 1}$ denote the effective sensing transmit weight. The received echo after matched processing is modeled as
\begin{equation}
\mathbf{r}=\beta \mathbf{a}(\boldsymbol{\eta})\mathbf{a}^T(\boldsymbol{\eta})\mathbf{u}+\mathbf{n},
\end{equation}
where $\beta$ is the complex reflection coefficient and $\mathbf{n}\sim\mathcal{CN}(\mathbf{0},\sigma_r^2\mathbf{I})$.

Define
\begin{equation}
\bm{\mu}(\boldsymbol{\eta}) \triangleq \mathbb{E}[\mathbf{r}]
= \beta \mathbf{a}(\boldsymbol{\eta})\mathbf{a}^T(\boldsymbol{\eta})\mathbf{u}.
\end{equation}
Since the observation is complex Gaussian with covariance $\sigma_r^2\mathbf{I}$, the Fisher information matrix (FIM) for $\boldsymbol{\eta}$ is
\begin{equation}
\mathbf{J}(\boldsymbol{\eta})
=
\frac{2}{\sigma_r^2}
\Re\left\{
\left(\frac{\partial \bm{\mu}}{\partial \boldsymbol{\eta}}\right)^H
\left(\frac{\partial \bm{\mu}}{\partial \boldsymbol{\eta}}\right)
\right\}.
\end{equation}
The position CRLB adopted in the main text is then obtained from the inverse FIM as
\begin{equation}
\mathrm{CRLB}(\boldsymbol{\eta})
=
\mathrm{tr}\!\left(\mathbf{J}^{-1}(\boldsymbol{\eta})\right),
\end{equation}
where a smaller value indicates better sensing accuracy.

\bibliographystyle{IEEEtran}
\bibliography{ref}

\end{document}